\documentstyle[12pt,psfig]{article}
\title{\bf Mass Loss on the Red Giant Branch and the 
       Second-Parameter Phenomenon}
\author{M.~Catelan \thanks{Hubble Fellow}\\
\vspace{1cm}\\
\normalsize University of Virginia,
            Department of Astronomy, USA}
\date{\mbox{}}
\begin{document}
\maketitle
\pagestyle{empty}
%
% WE REDEFINE THE plain LaTeX PAGESTYLE !!! 
% THIS PAGESTYLE WILL BE USED FOR THE FIRST PAGE ONLY !
%
\def\bull{\vrule height .9ex width .8ex depth -.1ex}
\makeatletter
\def\ps@plain{\let\@mkboth\gobbletwo
\def\@oddhead{}\def\@oddfoot{\hfil\tiny\bull\quad
``The Galactic Halo~: from Globular Clusters to Field Stars'';
35$^{\mbox{\rm th}}$ Li\`ege\ Int.\ Astroph.\ Coll., 1999\quad\bull}%
\def\@evenhead{}\let\@evenfoot\@oddfoot}
\makeatother
%
% AND DEFINE OUR MACROS FOR THE REFERENCE LIST
% I.E \beginrefer \refer and \endrefer
%
\def\beginrefer{\section*{References}%
\begin{quotation}\mbox{}\par}
\def\refer#1\par{{\setlength{\parindent}{-\leftmargin}\indent#1\par}}
\def\endrefer{\end{quotation}}
%
% BEGIN THE ABSTRACT CHAPTER WITH \noindent\small, ENCLOSE IT IN A GROUP
% AND BOLDFACE THE TITLE.
%
{\noindent\small{\bf Abstract:} 
The ``second-parameter effect" is characterized by the existence of globular 
clusters (GCs) with similar metallicity [Fe/H] (the ``first parameter") but 
different horizontal-branch (HB) morphologies. One of the primary 
second-parameter candidates is cluster age. In the present paper, we address 
the following issue: ``Are the age differences between second-parameter GCs, 
as derived from their main-sequence turnoff properties, consistent with 
their relative HB types?" In order to provide an answer to this question, 
several analytical formulae for the mass loss rate on the red giant branch 
are analyzed and employed. The case of M5 {\em vs.} Palomar~4/Eridanus is 
specifically discussed.  Our results show that, irrespective
of the mass loss formula employed, the relative turnoff ages of GCs are 
insufficient to explain the second-parameter phenomenon, unless GCs are 
younger than 10~Gyr. 
 }
%
% NOW COMES THE MAIN BODY OF THE ARTICLE
%
\section{Introduction}
Much recent debate has focused on the issue of whether age is the ``second 
parameter" of horizontal-branch (HB) morphology (the first parameter being 
metallicity [Fe/H]), or whether the phenomenon is instead much more complex, 
with several parameters playing an important role (VandenBerg 1999 and 
references therein). At the same time, most studies devoted toward 
this issue have adopted a {\em qualitative}, rather than {\em quantitative}, 
approach to the second-parameter phenomenon. More specifically, attempts to 
check whether a measured turnoff age difference between two GCs would be 
consistent with their relative HB types have been relatively rare. The main 
purpose of this paper is to report on some recent progress in this area.  

\section{Analytical Mass Loss Formulae for Cool Giants}
Mass loss on the red giant branch (RGB) is widely recognized as one of the 
most important ingredients, as far as the HB morphology goes (e.g., Catelan 
\& de Freitas Pacheco 1995; Lee et al. 1994; Rood et al. 1997). Up to 
now, investigations of the impact of RGB mass loss upon the HB morphology 
have mostly relied on Reimers' (1975) mass loss formula. We note, however, 
that Reimers' is by no means the only mass loss formula available for this 
type of study. In particular, alternative formulae have been presented by 
Mullan (1978), Goldberg (1979), and Judge \& Stencel (1991, hereafter JS91). 

\begin{figure}[t]
\centerline{\psfig{figure=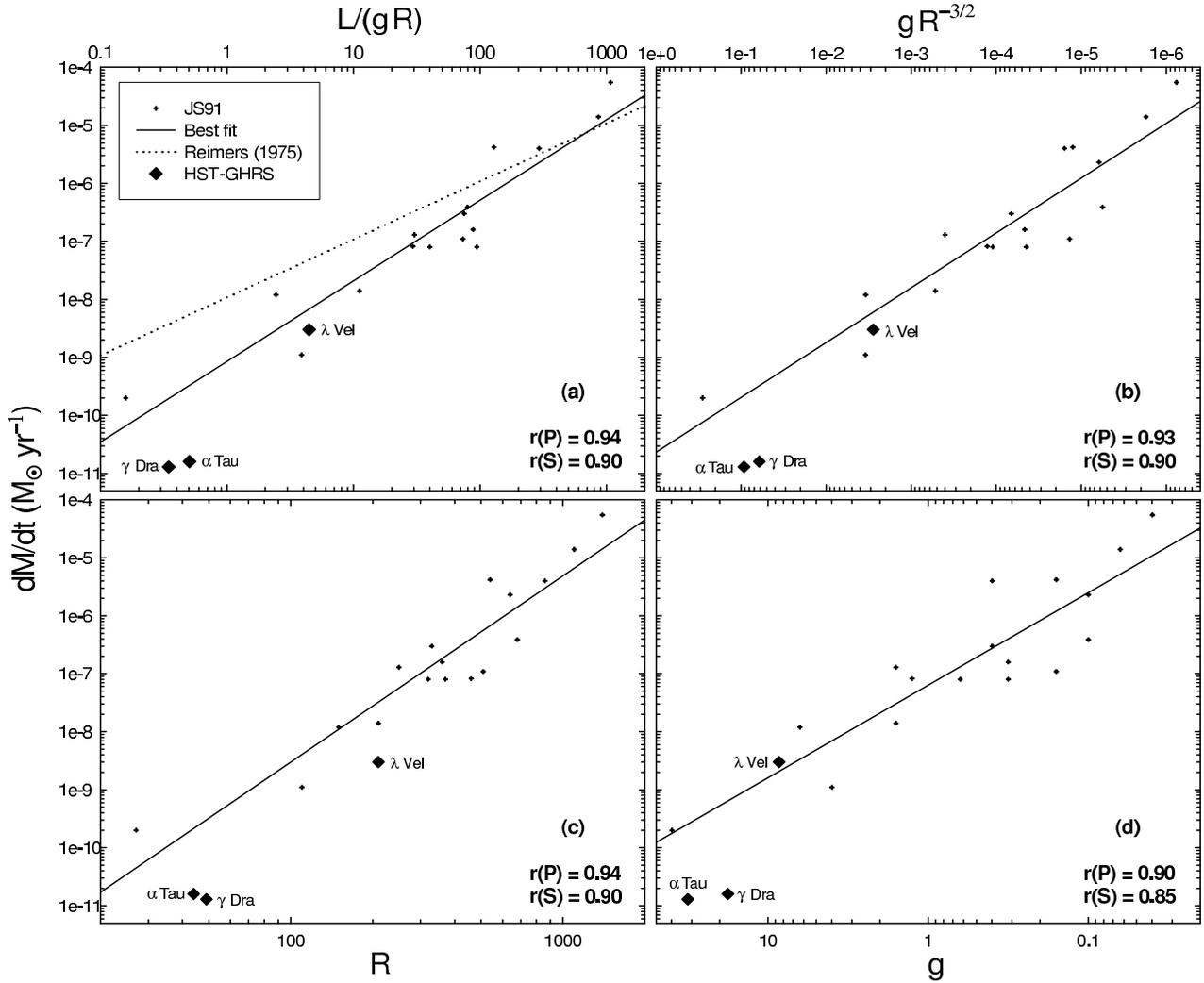,height=14cm,width=17cm}}
\caption{Mass loss rate for cool giant stars (JS91)
     is plotted against $L/(g\,R)$
     (panel a), $g/R^{\frac{3}{2}}$ (b), $R$ (c), 
     and $g$ (d). All quantities are given in solar units except 
     for gravity which is in cgs units. 
     The continuous lines represent least-squares fits to the data
     [eqs.~(1)--(4)]. The Pearson and Spearman correlation 
     coefficients are given. HST-GHRS results are also shown.}
\end{figure}

\subsection{Mass Loss Formulae Revisited}
As a first step in this project, we have undertaken a revision of all these 
formulae, employing the latest and most extensive dataset available in the 
literature---namely, that of JS91. The mass loss rates provided in JS91 
were compared against more recent data (e.g., Guilain \& Mauron 1996), 
and excellent agreement was found. If the distance adopted by JS91 lied 
more than about $2 \sigma$ away from that based on {\sc Hipparcos}
trigonometric parallaxes, the star was discarded. Only five stars turned 
out to be discrepant, in a sample containing more than 20 giants. 
Employing ordinary least-squares regressions, we find that the following 
formulae provide adequate fits to the data (see also Fig.~1):  

\begin{equation}
\frac{{\rm d}M}{{\rm d}t} = 8.5 \times 10^{-10} 
       \left(\!\!\!\begin{array}{c}
                L \\ \overline{g\,R}
            \end{array}\!\!\!\right)^{+1.4}\,\,
M_{\odot}\,{\rm yr}^{-1},
\end{equation}

\noindent with $g$ in cgs units, and $L$ and $R$ in solar units. As can  
be seen, this represents a ``generalized" form of Reimers' original mass 
loss formula, essentially reproducing a later result by Reimers (1987). 
The exponent (+1.4) differs from the one in Reimers' (1975) formula (+1.0) 
at $\approx 3\sigma$;  

\begin{equation}
\frac{{\rm d}M}{{\rm d}t} = 2.4 \times 10^{-11} 
       \left(\!\!\!\begin{array}{c}
                g \\ \overline{R^{\frac{3}{2}}}
            \end{array}\!\!\!\right)^{-0.9}\,\,
M_{\odot}\,{\rm yr}^{-1},
\end{equation}

\noindent likewise, but in the case of Mullan's (1978) formula; 

\begin{equation}
\frac{{\rm d}M}{{\rm d}t} = 1.2 \times 10^{-15}  \, R^{+3.2}\,\,
M_{\odot}\,{\rm yr}^{-1},
\end{equation}

\noindent idem, Goldberg's (1979) formula; 

\begin{equation}
\frac{{\rm d}M}{{\rm d}t} = 6.3 \times 10^{-8}  \, g^{-1.6}\,\,
M_{\odot}\,{\rm yr}^{-1},
\end{equation}

\noindent ibidem, JS91's formula. In addition, the expression 

\begin{equation}
\frac{{\rm d}M}{{\rm d}t} = 3.4 \times 10^{-12}  \, L^{+1.1}\, g^{-0.9}\,\,
M_{\odot}\,{\rm yr}^{-1}, 
\end{equation}

\noindent suggested to us by D.~VandenBerg, also provides a 
good fit to the data. ``Occam's razor" would favor equations~(3) or 
(4) in comparison with the others, but otherwise we are unable to 
identify any of them as being obviously superior. 

\subsection{Caveats}
We emphasize that mass loss formulae such as those given above should 
not be employed in astrophysical applications (stellar evolution, 
analysis of integrated galactic spectra, etc.) without keeping in 
mind these exceedingly important limitations: 

\begin{enumerate} 
\item As in Reimers' (1975) case, equations~(1) through 
      (5) were derived based on Population~I stars. Hence 
      they too are not well established for low-metallicity 
      stars. Moreover, there are only two first-ascent giants in 
      the adopted sample; 

\item Quoting Reimers (1977), ``besides the basic [stellar]
      parameters $\ldots$ the mass-loss process is probably also 
      influenced by the angular momentum, magnetic fields and close 
      companions. The {\em order of magnitude} of such effects is 
      completely unclear. Obviously, many observations will be
      necessary before we get a more detailed picture of stellar
      winds in red giants" (emphasis added). See also Dupree \&
      Reimers (1987); 

\item ``One should always bear in mind that a simple~$\ldots$ 
      formula like that proposed can be expected to yield only 
      correct order-of-magnitude results if extrapolated to the 
      short-lived evolutionary phases near the tips of the giant 
      branches" (Kudritzki \& Reimers 1978);

\item ``Most observations have been interpreted using models that 
      are relatively simple (stationary, polytropic, spherically 
      symmetric, homogeneous) and thus `observed' mass loss rates 
      or limits may be in error by orders of magnitude in some 
      cases" (Willson 1999);  

\item The two first-ascent giants analyzed by 
      Robinson et al. (1998) using HST-GHRS, $\alpha$~Tau and 
      $\gamma$~Dra, appear to both lie about one order of magnitude 
      below the relations that best fit the JS91 data---two orders 
      of magnitude in fact, if compared to Reimers' formula (see 
      Fig.~1). The K supergiant $\lambda$~Vel, analyzed by the same 
      group (Mullan et al. 1998), appears in much better agreement  
      with the adopted dataset and best fitting relations. 
\end{enumerate}

{\em In effect, mass loss on the RGB is an excellent, but virtually 
untested, second-parameter candidate.} It may be connected to GC  
density, rotational velocities, and abundance anomalies on the RGB. 
It will be extremely important to study mass loss in first-ascent, 
low-metallicity giants---{\em in the field and in GCs alike}---using 
the most adequate ground- and space-based facilities available, or 
expected to become available, in the course of the next decade. 
Moreover, in order to properly determine how (mean) mass loss 
behaves as a function of the fundamental physical parameters and 
metallicity, astrometric missions much more accurate than 
{\sc Hipparcos}, such as SIM and GAIA, will certainly be necessary.  

In the meantime, we suggest that using several different mass loss 
formulae (such as those provided in Sect.~2.1) constitutes a better 
approach than relying on a single one. This is the approach that we 
are going to follow in the rest of this paper. 

\begin{figure}[t]
\centerline{\psfig{figure=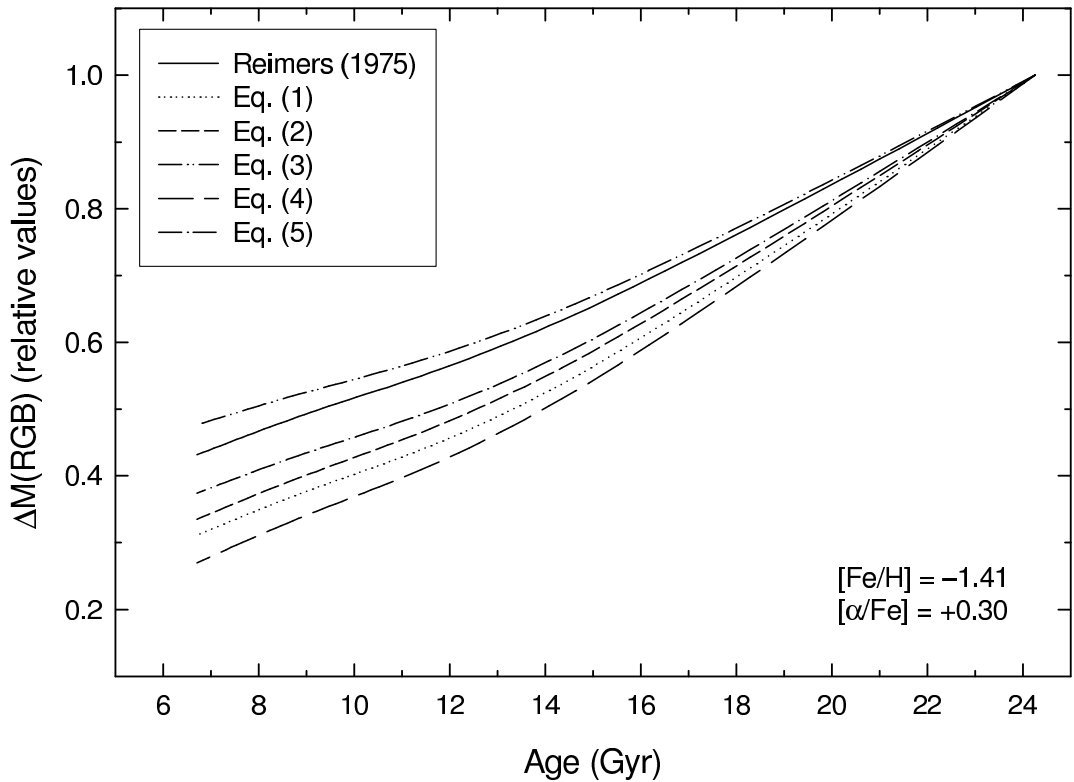}}
\caption{Dependence of mass loss on age, for the chemical composition 
indicated at the lower right-hand corner. For each mass loss formula, 
mass loss values were normalized to the highest value, attained at an
age of 24.3~Gyr.}
\end{figure}

\section{Implications for the Amount of Mass Lost by First-Ascent Giants} 
The latest RGB evolutionary tracks by VandenBerg et al. (2000) were 
employed in an investigation of the amount of mass lost on the RGB and 
its dependence on age. The effects of mass loss upon RGB evolution 
were ignored. In Figure~2, the mass loss--age relationship is 
shown for each of equations~(1) through (5), and also for Reimers' 
(1975) formula, for a metallicity ${\rm [Fe/H]} = -1.41$, 
$[\alpha/{\rm Fe}] = +0.30$. Even though the formulae from Section~2.1
are all based on the very same dataset, the implications differ from 
case to case. 

\begin{figure}[t]
\centerline{\psfig{figure=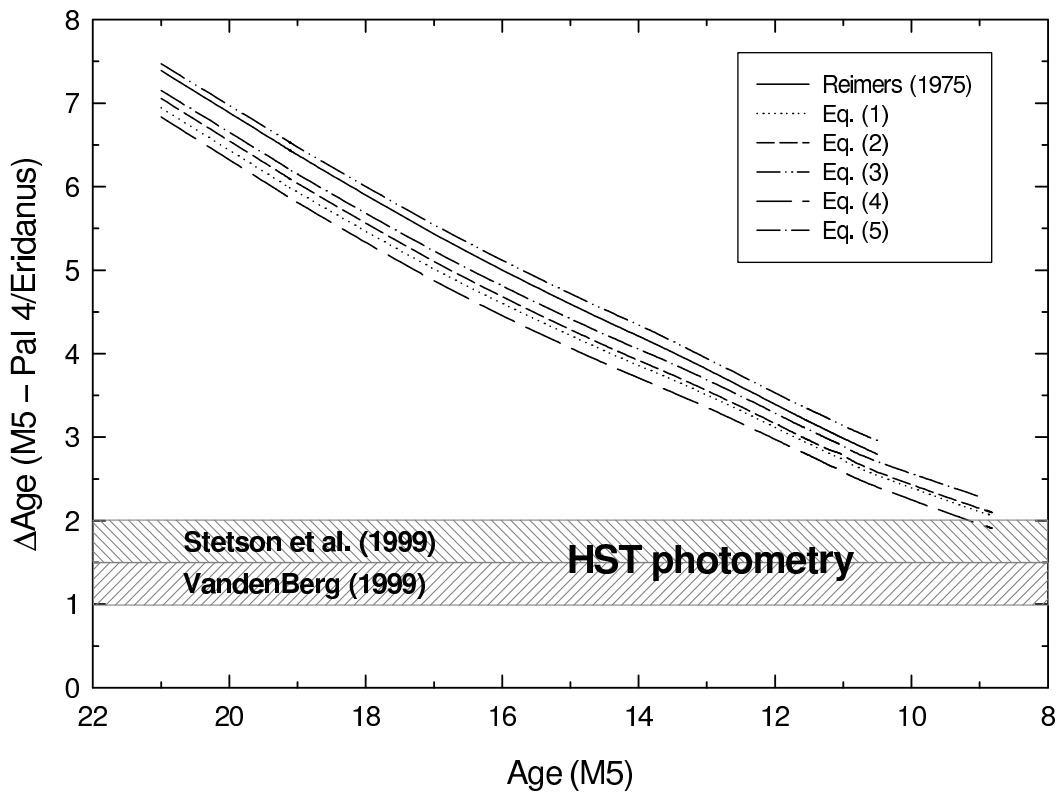}}
\caption{Difference in age (in Gyr) between M5 and Pal~4/Eridanus, 
derived for the several indicated mass loss formulae, as a function 
of the assumed M5 age (also in Gyr). The hatched areas correspond to 
the range in turnoff age differences between M5 and Pal~4/Eridanus, 
as estimated by Stetson et al. (1999) and VandenBerg (1999) from 
deep HST photometry.}
\end{figure}

\section{The Second-Parameter Effect: the Case of Pal~4/Eridanus {\em vs.} 
         M5} 
Stetson et al. (1999) presented $V$,~$V-I$ color-magnitude diagrams for 
the outer-halo GCs Pal~4 and Eridanus, based on HST images. Analyzing 
their {\em turnoff} ages, they concluded that Pal~4 and Eridanus are 
younger than M5, a GC with similar metallicity but a bluer HB, by 
$1.5-2$~Gyr. Based on the same data, VandenBerg (1999) obtained a 
smaller age difference: $1-1.5$~Gyr. {\em Are these values 
consistent with the relative HB types of Pal~4/Eridanus vs. M5?} 

To answer this question, we constructed detailed synthetic HB 
models (based on the evolutionary tracks described in Catelan et 
al. 1998) for M5 and Pal~4/Eridanus, thus obtaining the difference 
in mean HB mass between them---which was then transformed to a 
difference in age with the aid of the mass loss formulae from 
Section~2 and the RGB mass loss results from Section~3. Figure~3 
shows the age difference thus obtained as a function of the adopted 
M5 age, in comparison with the turnoff determinations from Stetson 
et al. (1999) and VandenBerg (1999). Our assumed reddening values 
for Pal~4/Eridanus come from Schlegel et al. (1998); had the 
Harris (1996) values been adopted instead, the curves in Figure~3 
corresponding to the different mass loss formulae would all be 
shifted upwards.

\section{Conclusions}
As one can see from Figure~3 our results indicate that, irrespective 
of the mass loss formula employed, age cannot be the only ``second 
parameter" at play in the case of M5 {\em vs.} Pal~4/Eridanus, unless 
these GCs are younger than 10~Gyr.

%
% USE A SECTION WITHOUT NUMBER FOR THE ACKNOWLEDGEMENTS
%
\section*{Acknowledgements}
The author wishes to express his gratitude to D.A.~VandenBerg for 
providing many useful comments and suggestions, and also for making 
his latest evolutionary sequences available in advance of publication. 
Support for this work was provided by NASA through Hubble Fellowship 
grant HF--01105.01--98A awarded by the Space Telescope Science 
Institute, which is operated by the Association of Universities for 
Research in Astronomy, Inc., for NASA under contract NAS~5--26555. 
%
% BEGIN THE REFERENCE LIST WITH \beginrefer
% USE \refer BEFORE THE REFERENCES AND BEGIN A NEW PARAGRAPH AFTER THE 
% REFERENCE !
% DO NOT FORGET TO END THE LIST WITH \endrefer
%
 
\beginrefer
\refer Catelan M., Borissova J., Sweigart A.V., Spassova N., 1998, 
ApJ 494, 265

\refer Catelan M., de Freitas Pacheco J.A., 1995, A\&A 297, 345 

\refer Dupree A.K., Reimers D., 1987. In: Kondo Y., et al. (eds.) 
Exploring the Universe with the IUE Satellite. Dordrecht, Reidel, 
p.~321

\refer Goldberg L., 1979, QJRAS 20, 361

\refer Guilain Ch., Mauron N., 1996, A\&A 314, 585 

\refer Harris W.E., 1996, AJ 112, 1487

\refer Judge P.G., Stencel R.E., 1991, ApJ 371, 357 (JS91) 

\refer Kudritzki R.P., Reimers D., 1978, A\&A 70, 227

\refer Lee Y.-W., Demarque P., Zinn R., 1994, ApJ 423, 248

\refer Mullan D.J., 1978, ApJ 226, 151

\refer Mullan D.J., Carpenter K.G., Robinson R.D., 1998, ApJ 495, 
927

\refer Reimers D., 1975. In: M\'emoires de la Societ\'e Royale 
des Sciences de Li\`ege, 6e serie, tome VIII, 
Probl\`emes D'Hydrodynamique Stellaire, p.~369

\refer Reimers D., 1977, A\&A 57, 395

\refer Reimers D., 1987. In: Appenzeller I., Jordan C. (eds.) IAU 
Symp. 122, Circumstellar Matter. Dordrecht, Kluwer, p.~307  

\refer Robinson R.D., Carpenter K.G., Brown A., 1998, ApJ 503, 396

\refer Rood R.T., Whitney J., D'Cruz N., 1997. In: Rood R.T., 
Renzini A. (eds.) Advances in Stellar Evolution. Cambridge, 
Cambridge University Press, p.~74 

\refer Schlegel D.J., Finkbeiner D.P., Davis M., 1998, ApJ 500, 525

\refer VandenBerg D.A., 1999, ApJ, submitted  

\refer VandenBerg D.A., Swenson F.J., Rogers F.J., Iglesias 
C.A., Alexander D.R., 2000, ApJ 528, in press (January $20^{\rm th}$ 
issue) 

\refer Willson L.A., 1999. In: Livio M. (ed.) Unsolved Problems in 
Stellar Evolution. Cambridge, Cambridge University Press, p.~227

\endrefer           
\end{document}